\documentclass[notitlepage,floatfix,aps,pra,twocolumn,amsmath,amssymb,superscriptaddress,]{revtex4-2}
\usepackage[pretty,uselistings,final]{revquantum}
\usepackage{graphicx}

\usepackage{comment}
\usepackage{xcolor}

\definecolor{forestgreen}{rgb}{0.13, 0.55, 0.13}

\begin{document}

\title{A Bright Source of Telecom Single Photons Based on Quantum Frequency Conversion}

\author{Christopher L. Morrison}
\email{clm6@hw.ac.uk}

\author{Markus Rambach}
\altaffiliation[Present address: ]{Centre for Engineered Quantum Systems, School of Mathematics and Physics, University of Queensland, QLD 4072, Australia}

\author{Zhe Xian Koong}

\author{Francesco Graffitti}

\author{Fiona Thorburn}

\author{Ajoy K. Kar}

\affiliation{Institute of Photonics and Quantum Sciences, School of Engineering and Physical Sciences, Heriot-Watt University, Edinburgh EH14 4AS, UK}

\author{Yong Ma}

\affiliation{College of Optoelectronic Engineering, Chongqing University of Posts and Telecommunications, Chongqing 400065, China}

\author{Suk-In Park}

\author{Jin Dong Song}

\affiliation{Center for Opto-Electronic Materials and Devices Research, Korea Institute of Science and Technology, Seoul 02792, Republic of Korea}

\author{Nick G. Stoltz}

\affiliation{Materials Department, University of California, Santa Barbara, California 93106, USA}

\author{Dirk Bouwmeester}
\affiliation{Huygens-Kamerlingh Onnes Laboratory, Leiden University, P.O. Box 9504, 2300 RA Leiden, Netherlands}
\affiliation{Department of Physics, University of California, Santa Barbara, California 93106, USA}

\author{Alessandro Fedrizzi}

\author{Brian D. Gerardot}
 
 \affiliation{Institute of Photonics and Quantum Sciences, School of Engineering and Physical Sciences, Heriot-Watt University, Edinburgh EH14 4AS, UK}

\date{\today}

\begin{abstract}
On-demand indistinguishable single photon sources are essential for quantum networking and communication. Semiconductor quantum dots are among the most promising candidates, but their typical emission wavelength renders them unsuitable for use in fibre networks. 
Here, we present quantum frequency conversion of near-infrared photons from a bright quantum dot to the telecommunication C-band, allowing integration with existing fibre architectures. 
We use a custom-built, tunable 2400~nm seed laser to convert single photons from 942~nm to 1550~nm in a difference frequency generation process. 
We achieve an end-to-end conversion efficiency of $\sim$35\%, demonstrate count rates approaching 1 MHz at 1550~nm with $g^{\left(2\right)}\left(0\right) = 0.04$, and achieve Hong-Ou-Mandel visibilities of 60\%.
We expect this scheme to be preferable to quantum dot sources directly emitting at telecom wavelengths for fibre based quantum networking.
\end{abstract}

\maketitle 

Semiconductor quantum dots (QDs) are a leading technology for bright, indistinguishable single-photon sources. 
QDs emitting in the 920 nm - 980 nm window have been used as a source of single photons with count rates upwards of 10~MHz, indistinguishability greater than 95\%, and low multi-photon contributions on the order of 1\%~\cite{tomm_bright_2020,wang_towards_2019-1,somaschi_near-optimal_2016}.
Yet, in order to be compatible with existing telecommunication technology, an ideal single-photon source should operate in the telecommunication C-band, around 1550~nm, where fibre loss is minimal.
While there has been recent progress in producing QD sources that directly emit single photons in the C-band~\cite{olbrich_polarization-entangled_2017,anderson_quantum_2020,2019arXiv191204782Z}, achieving high-quality single-photon emission at high rates remains an open challenge, with the best-performing single photon sources to date being confined to near-infrared (NIR)~\cite{wang_boson_2019} wavelengths.

One route to bridge the gap to the C-band is quantum frequency conversion (QFC), converting single photons from a NIR QD to telecommunication wavelengths.
Quantum frequency conversion is a nonlinear process where a single-photon input is mixed with a strong seed beam producing a single-photon output at either the sum or difference frequency.
QFC can in principle be noise free and therefore preserve the quantum statistics of near-infrared emitters. 
Frequency conversion of QD sources has been demonstrated from 700~nm to the telecommunication O-band~\cite{zaske_visible--telecom_2012}, and from 900~nm to the C-band \cite{Kambs:16,PhysRevB.97.195414}, culminating in the remote two-photon interference between independent downconverted QD sources~\cite{weber_two-photon_2019}. 
Spin-photon entanglement between a 910~nm QD and a 1560-nm photon has been demonstrated using frequency conversion seeded by ultrafast pulses~\cite{de_greve_quantum-dot_2012,pelc_downconversion_2012}.
QFC of QDs has also been demonstrated in nano-photonic circuits using four-wave mixing in silicon nitride~\cite{singh_quantum_2019}.
However, a source that is simultaneously bright, pure and coherent has not been demonstrated in QDs emitting directly at telecom wavelengths. Here we demonstrate a frequency-converted InGaAs quantum dot source approaching 1 MHz count rates at 1550 nm, with $g^{\left(2\right)}\left(0\right)$ around 4\% and HOM visibilities of 60\%. 

\begin{figure*}[tb]
\centering\includegraphics[width=1\textwidth]{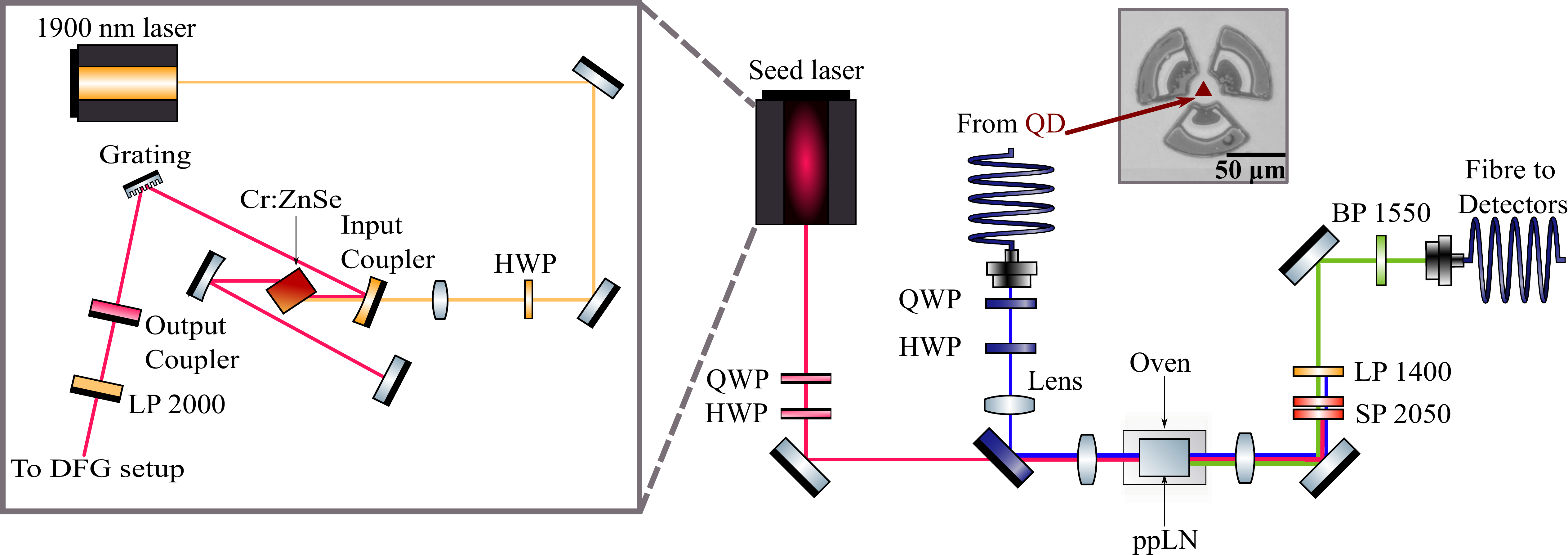}
\caption{
Difference-frequency generation schematic. 
Blue lines indicate the optical path of the QD photons. 
Pink lines represent the path of the 2401~nm seed light.
The polarisation of both beams is aligned to the extraordinary axis of the ppLN crystal with a quarter-wave plate (QWP) and half-wave plate (HWP).
A 100 mm focal length lens is used in the QD beam path to mode match the seed beam at the waveguide facet.
The green lines represent the converted 1550~nm light after the frequency conversion.
The converted light is sent through two short pass filters at 2050~nm (SP 2050), a 1400~nm long pass filter (LP 1400) and a bandpass filter at 1550~nm (BP 1550) before being collected in a single mode fibre for detection.
The inset shows the experimental layout used to produce the 2401~nm laser light for QFC. 
The seed laser is pumped by a commercial Thulium fibre laser (yellow lines).
The pump beam polarisation is prepared with a HWP to reduce loss in the laser cavity due to Fresnel reflections.
The light enters the cavity through a partially reflective curved mirror which acts as the input coupler at 1900~nm and a focusing mirror at 2401~nm.
A longpass filter with a cutoff wavelength of 2000~nm (LP 2000) is placed after the output coupler to remove unabsorbed 1900~nm pump light.
}
\label{DFG_setup}
\end{figure*}

\label{experimental setup}
For difference frequency generation, energy conservation demands $\left(1/\lambda_{\text{in}} -1/\lambda_{\text{seed}}\right)^{-1}=\lambda_{\text{out}}$. 
Our 942~nm InGaAs QD source requires a seed wavelength of 2401~nm to generate output photons at 1550~nm. 
The seed beam is produced in a chromium doped zinc selenide (Cr:ZnSe) laser while the difference frequency generation occurs in a periodically poled lithium niobate (ppLN) waveguide.

\label{dot section}
Near-infrared photons are generated by a single self-assembled InGaAs/GaAs QD coupled to a high quality ($Q\approx 4.4\times 10^4$) oxide-apertured micropillar cavity, as shown in Fig.~\ref{DFG_setup}.
The QDs are embedded in a \textit{p-i-n} diode structure~\cite{2008PhDT.......129R,rakher_externally_2009} which enables charge control and tuning of the QD emission to the cavity mode via the quantum-confined Stark effect.
The sample is kept at a temperature of $4\,\mathrm{K}$ in a closed-cycle helium flow cryostat.
A dark-field confocal microscope is used to excite and collect the scattered photons from the QD before filtering in a cross-polarisation scheme with a $\sim 10^7$ extinction ratio to suppress the excitation laser.

Pulsed excitation of the QD is performed using a mode-locked titanium:sapphire laser with a repetition rate of 80.3~MHz and pulse duration of 10~ps.
The QD output is then detected by superconducting nanowire single photon detectors (SNSPDs), with a nominal quantum efficiency of $\sim 90\,\%$ at 950~nm. 
We exploit a neutral exciton QD transition ($X^0$), resonantly coupled to the cavity with a Purcell factor of $\sim 4$ and an emission wavelength of 942.33~nm.
The $T_2$ coherence time of the emission, measured using standard Fourier spectroscopy under $\pi$-pulse resonant excitation, shows $T_2=0.348\,(2)\,\mathrm{ns}$, corresponding to a linewidth of $915\,(5)\, \mathrm{MHz}$.
This value is $\approx 1.5$ times larger than the transform-limited linewidth ($h/T_1=607\, \mathrm{MHz}$), with an independently measured lifetime of $T_1=0.2622\,(1)\,\mathrm{ns}$. 

The seed laser for the QFC stage consists of a z-cavity resonator with a Cr:ZnSe crystal, a gain medium with an emission spectrum spanning 1900-3300~nm~\cite{macdonald_ultrabroad_2015,wagner_continuous-wave_1999,sorokina_broadly_2002}, see inset in Fig.~\ref{DFG_setup}.
This laser design allows for both continuous-wave ~\cite{wagner_single-frequency_2004} and mode-locked operation generating pulses as short as 43~fs~\cite{vasilyev_high_2014}. 
Here, we operate with a narrowband CW seed to drive the QFC. Our 2401~nm laser system is pumped by a thulium-doped fibre laser (IPG Photonics TLR-20-LP) which has a maximum CW output power of 20~W at 1900~nm. The pump light is focused into the Cr:ZnSe crystal using a 100~mm $\text{CaF}_{2}$ lens. The cavity consists of a dichroic input coupler (50~mm radius of curvature (ROC), transmissive at 1900~nm reflective at 2400~nm), a gold mirror (50~mm ROC), two plane silver mirrors and the Cr:ZnSe crystal. The crystal is placed at Brewster angle to minimise losses due to reflection.
A diffraction grating (450 lines/mm) is inserted into the cavity to control the emission wavelength. 
The output coupler (Layertec) has a transmission of 60\% at 2401~nm, allowing a good trade-off between the cavity enhancement and available output power.

Fig.~\ref{DFG_setup} shows our difference-frequency generation (DFG) setup.
The DFG, one special case of QFC, takes place in a 48~mm periodically-poled lithium niobate crystal (ppLN, NTT Electronics).
The chip contains multiple ridge waveguides with poling periods ranging from 26.00~$\mu$m to 26.25~$\mu$m.
These poling periods are designed for type-0 DFG from 942~nm to 1550~nm.
Quarter- and half-wave plates are used to align the polarisation of the incoming single photons at 942~nm and the generated pump light at 2401~nm to the extraordinary axis of the crystal.
Seed light from the laser is overlapped with single photons from the QD using a dichroic mirror (Omega Optical) and coupled into the waveguide using a NIR coated aspheric lens with a focal length of 11~mm.
A NIR coated lens is used to match the beam size of the single photons to the seed beam and to compensate for the chromatic aberration of the aspheric lenses.
The converted 1550~nm light is collimated with an 11~mm NIR-coated aspheric lens and sent towards a filtering stage. 

The filtering stage consists of two shortpass filters at 2050~nm ($>$OD 4), which are used to remove seed light impinging on the collection fibre; a longpass filter at 1400~nm ($>$OD 5) to remove weakly phase-matched second-harmonic generation from the seed beam and unconverted quantum dot light; and finally, a 2.8~nm full-width-at-half-maximum (FWHM) bandpass filter ($>$OD 4, BP 1550) to isolate the converted single photons. 
The converted 1550~nm light is collected into a single mode fibre with 86\% coupling efficiency and sent to SNSPDs with a nominal quantum efficiency greater than 80\%.

\label{experimental results}

The DFG conversion efficiency is characterised by sending CW coherent light from a 942~nm laser (Toptica DL Pro) into the QFC setup. 
For CW-seeded QFC, the conversion efficiency is almost independent of the temporal mode of the input light~\cite{brecht_quantum_2011}. 
This allows characterisation with a CW beam despite the single photons' decaying exponential wavepacket. 
In the case that we can treat the seed beam as unamplified, QFC is expressed as a beam-splitter Hamiltonian between two different frequency modes~\cite{kumar_quantum_1990}. 
This means the measured conversion efficiency is independent of the input intensity.
These two factors ensure the conversion efficiency measured with a low power $\left(500 \text{ $\mu$W}\right)$ CW coherent field is equivalent to the single-photon conversion efficiency. 

\begin{figure}[tb]
\centering\includegraphics[width=1\columnwidth]{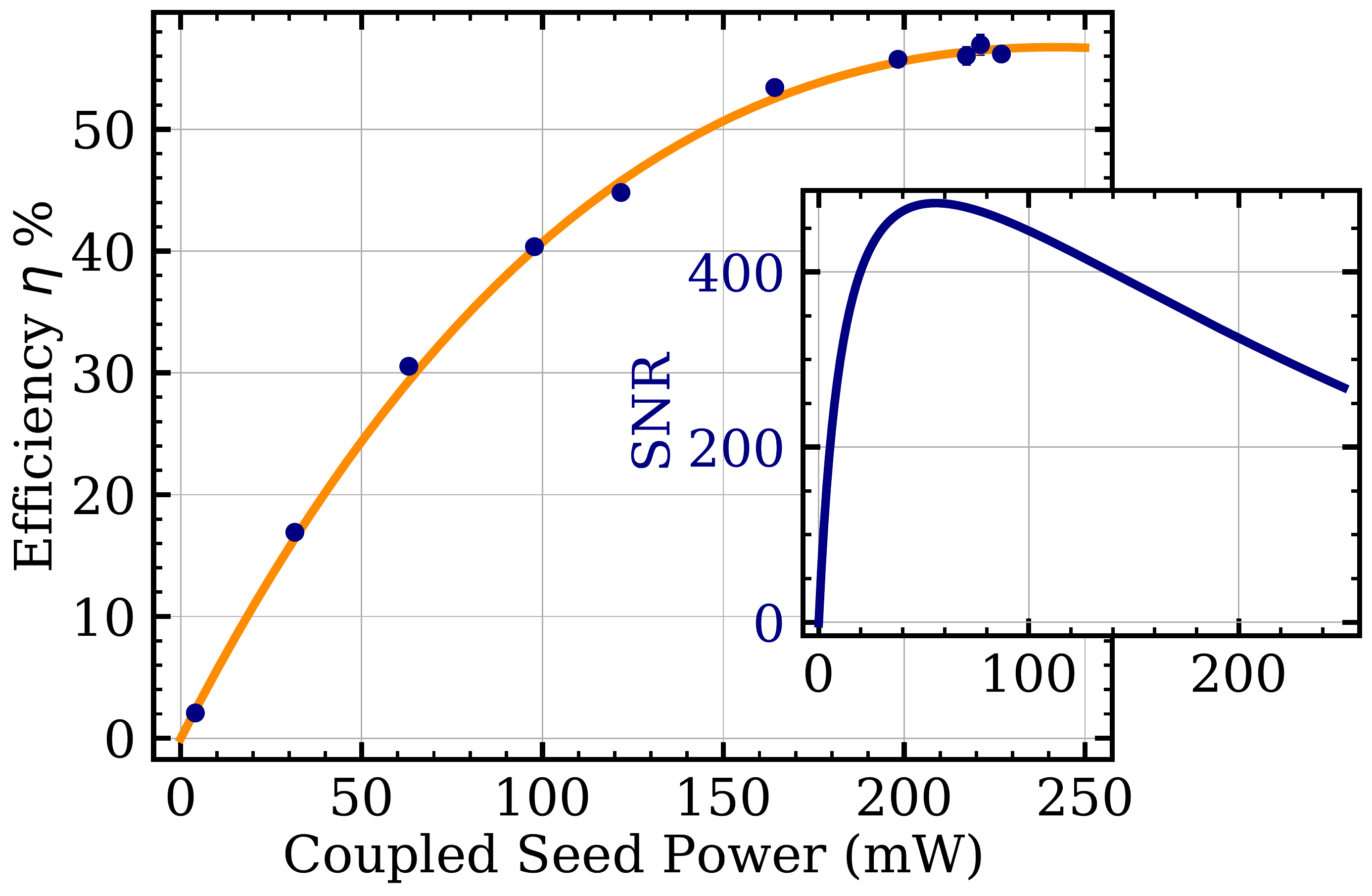}
\caption{Conversion efficiency of the difference frequency generation process as a function of seed power coupled into the waveguide. 
The power in the waveguide is determined by measuring the pump power after the waveguide and factoring out the loss through the c-coated aspheric collimation lens. 
The transmission through this lens is measured to be 64\,\% at the pump wavelength.
The data is fitted with Eq.~\ref{conversion_efficiency}. 
Inset shows the signal to noise ratio for off-resonant excitation with a measured noise count rate of $12\,(1)$~Hz/mW.
\label{conversion_efficiency_plot}}
\end{figure}

\begin{figure}[b]
\centering\includegraphics[width=1\columnwidth]{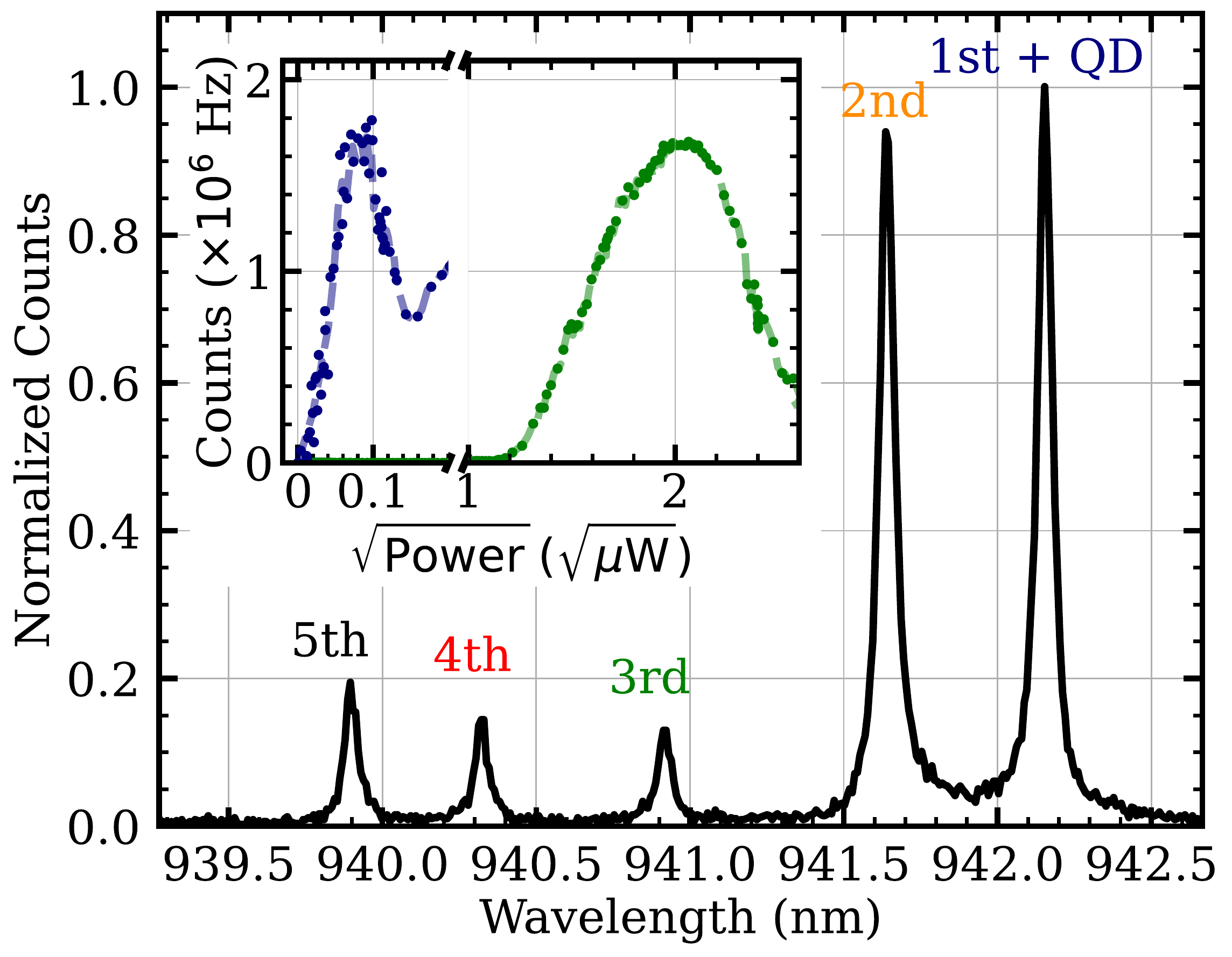}
\caption{ Photo-luminescence emission profile of the QD under 820~nm excitation shows cavity modes up to the $5^\mathrm{th}$ order. 
The QD is resonantly coupled to the 1st cavity mode.
Inset shows the detected count rate as a function of the excitation power when the excitation laser is resonant to the 1st (resonant, blue) or 3rd (non-resonant, green) cavity mode. 
\label{qd_characterisation}}
\end{figure}

\begin{figure*}[tb]
\centering\includegraphics[width=1\textwidth]{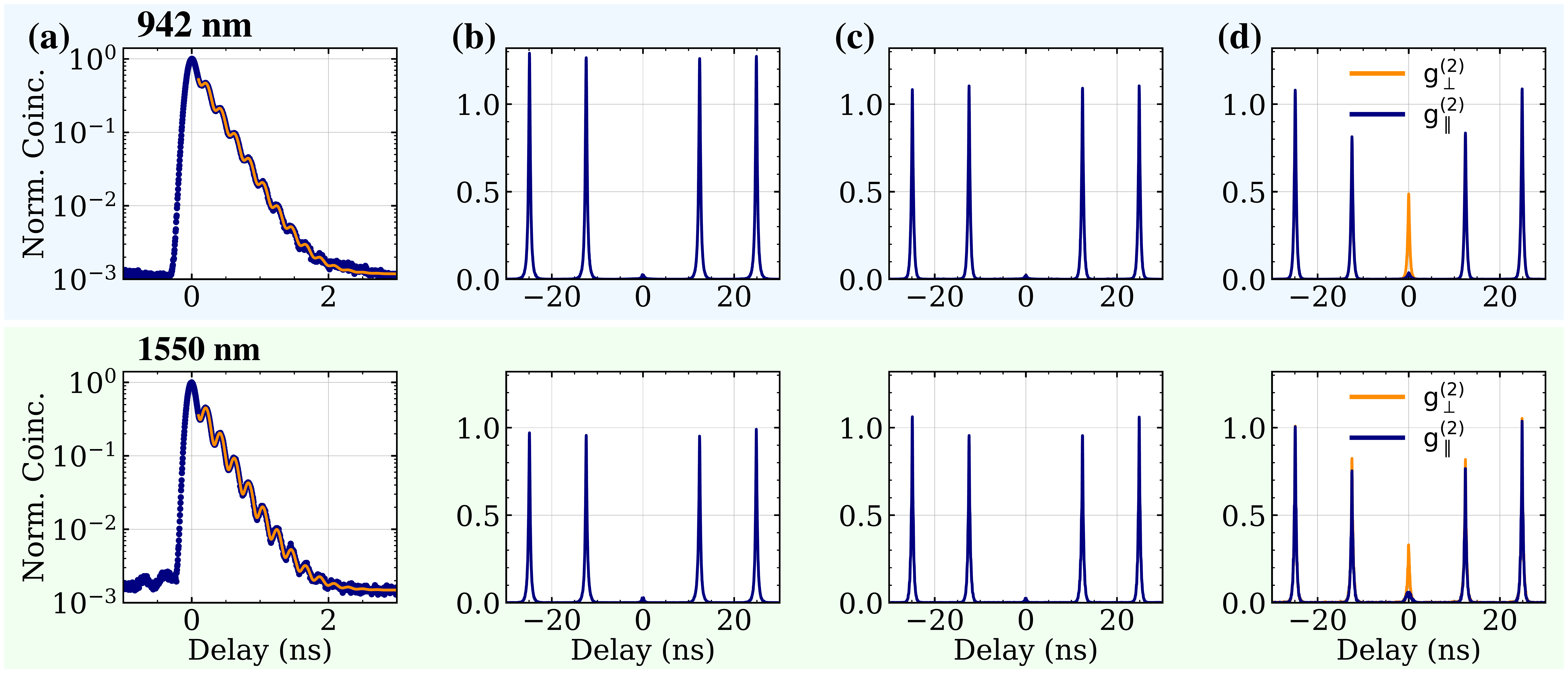}
\caption{
Characterisation of the single photon properties before (upper row) and after (lower row) QFC.
\textbf{(a)} Time-resolved emission spectra under pulsed resonant excitation reveals an exponential decay which gives the emitter's lifetime $T_1$ and a fast oscillation indicating the quantum beating between the fine-structure peaks of the neutral exciton emission, $\Delta_{\rm fss} = 4.807\,(3)\,\mathrm{GHz}$.
\textbf{(b, c)} Second-order intensity correlation histogram $g^{(2)}$, of the emitted photons under off-resonant (b) and resonant (c) excitation. 
The lack of coincidences in the central peak indicates the low probability of multi-photon emission.
\textbf{(d)}  Two-photon interference of consecutively scattered photons delayed by $12.5\,\mathrm{ns}$, prepared in cross ($g^{(2}_\perp$) and parallel ($g^{(2}_\parallel$) polarisations, under resonant $\pi$-pulse excitation. The extracted photon indistinguishability, given by ratio of zero-delay-coincidences from both configurations, along with the extracted values ($T_1$ and $g^{(2)}(0)$) from the fits (solid lines) are summarised in Table.~\ref{tab:table1}.
}
\label{fig:qd_fig_2}
\end{figure*}

Figure~\ref{conversion_efficiency_plot} shows the internal conversion efficiency for 942~nm light, measured by comparing output 1550~nm light to the 942~nm coupled through the waveguide with the seed laser blocked. 
This factors out coupling losses into the waveguide which are measured to be $17\,\%$.
The data is fitted with~\cite{roussev_periodically_2004}
\begin{equation}
\eta = \eta_{max} \sin ^2 \left(\sqrt{\eta_{nor}P}~L\right), 
\label{conversion_efficiency}
\end{equation}
where $\eta_{max}$ is the maximum possible conversion efficiency, $\eta_{nor}$ is the normalised conversion efficiency of the process, $P$ is the input power, and $L$ is the waveguide length. 
The fit gives a normalised conversion efficiency (to waveguide length in the limit of small pump powers~\cite{roussev_periodically_2004}) is $\eta_{nor} = 44\left(1\right)\, \% \text{ /} \left( \text{W cm}^2 \right)$. 
The maximal external conversion efficiency, the ratio of photons collected into single-mode fibre after the conversion stage to the number of NIR photons impinging on the waveguide, $\eta_{max} = 38\,\%$, leading to a maximum internal conversion efficiency of $56.7\left(4\right)\,\%$ when taking losses into account.
This $\eta_{max}$ is higher than previously reported values for NIR QD frequency conversion to 1550~nm with similar waveguides~\cite{weber_two-photon_2019}.
We would like to highlight that we achieve SNRs~$>250$ for all seed powers (inset Fig.~\ref{conversion_efficiency_plot}), meaning that the noise contribution of the DFG process towards the converted single photons is minimal.

We now compare the characteristics of the converted telecom photons to the QD NIR photons. 
We tune the QD into resonance with the first cavity mode, and excite either resonantly or non-resonantly into the third cavity mode; see Fig.~\ref{qd_characterisation} for spectral properties of the cavity under 820 nm excitation. 
The inset in Fig.~\ref{qd_characterisation} shows the detected count rates for these two excitation scenarios as a function of power: Rabi oscillations are observed for resonant excitation while a clear maximum is observed for non-resonant excitation. 
For non-resonant characterisation, the QD photons are spectrally filtered with a grating filter with a 30~GHz FWHM to suppress the excitation laser. We detect a count rate of $1.85(5)~\mathrm{MHz}$ at an excitation power of $6.8\,\mu\mathrm{W}$. The grating filter was removed when characterising the converted photons as low-loss bandpass filters were used at 1550~nm. For resonant driving, we optimize the excitation power to the $\pi$-pulse and detect a count rate of $ 1.46(4)~\mathrm{MHz}$.
This value is slightly lower than for off-resonant excitation due to the presence of spectral fluctuations~\cite{Santana_generating_2017}.
After QFC, the detected count rate at 1550~nm, for the off-resonant and on-resonant case is $856(18)~\mathrm{kHz}$ and $ 456(14)~\mathrm{kHz}$, respectively. Comparing the NIR and telecom counts under resonant excitation gives an end-to-end conversion efficiency of $\approx 35\,\%$, after accounting for the difference in the detection efficiency of both NIR ($\sim 90\%$) and telecom C-band ($\sim 80\%$) detectors. 
This agrees well with the measured loss budget through the optical components including the conversion efficiency. The difference in efficiency for off-resonant excitation is accounted for by the loss of the grating filter.

Figure~\ref{fig:qd_fig_2} shows the comparison between the performance of the QD signal before and after QFC.
The lifetime measured under resonant excitation in Fig.~\ref{fig:qd_fig_2}(a) remains unchanged within experimental error after conversion.
The oscillation in the time-resolved emission, indicative of the quantum beating of the $X^0$ fine-structure splitting, shows a frequency of $4.807\,(3)\,\mathrm{GHz}$.
The equivalent oscillation after the QFC process is unchanged ($4.803\,(1)\,\mathrm{GHz}$), indicating that the CW-seeded frequency conversion preserves the temporal mode of the input photons. 

Next, we measure the second-order intensity correlation $g^{(2)}$ using a Hanbury-Brown and Twiss (HBT) interferometer. 
For a perfect single-photon source $g^{(2)}(0)=0$, indicating the absence of multi-photon emissions.
Under off-resonant driving, Fig.~\ref{fig:qd_fig_2}(b), we observe a slight increase from $g^{(2)}(0)=0.045\,(0)$ to $g^{(2)}(0)=0.051\,(1)$ before and after the QFC process, respectively.
We observe similar values under resonant driving, Fig.~\ref{fig:qd_fig_2}(c), demonstrating near-ideal single-photon emission with $g^{(2)}(0)=0.040\,(0)$ and $g^{(2)}(0)=0.043\,(1)$ before and after the QFC process, respectively.
The slight increase in the normalized coincidences in the uncorrelated side peaks in the HBT histogram is due to blinking of the emitters, a common effect resulting from QD coupling to the solid-state charge environment~\cite{davanco_multiple_2014}.
The imperfection in $g^{(2)}(0)$ can be due to imperfect suppression of the cavity emission due to cavity feeding~\cite{PhysRevB.77.161303,PhysRevLett.103.207403,PhysRevLett.103.027401}, slight imperfection in the wave-plate retarders used in our confocal microscope, and presence of multi-photon capture processes~\cite{peter_fast_2007,lanco_highly_2015}.
Nevertheless, with a modest increase in $g^{(2)}(0)$ after the QFC process, we have demonstrated near background-free single photon frequency conversion from NIR to telecom C-band, with the photon number purity predominately limited by the quantum dot.

To demonstrate that our QFC setup preserves photon coherence, we perform Hong-Ou-Mandel (HOM) interference between photons emitted from two consecutive excitation pulses.
We use an unbalanced Mach-Zehnder interferometer with a delay of 12.5~ns to match photons temporally on a 50/50 beam splitter. 
We measure the coincidence counts for parallel and perpendicular polarised photons and evaluate the visibility as $\mathrm{V_{HOM}}=1-g^{(2)}_\parallel/g^{(2)}_\perp$. 
For a pair of indistinguishable photons, $\mathrm{V_{HOM}}=1$.
For resonant excitation we achieve an interference visibility of $\mathrm{V_{HOM}}=0.88\,(1)$ before QFC. We calculate the single photon indistinguishability $ M_s$ as $ M_s = (\mathrm{V_{HOM}}+g^{(2)}(0))/(1-g^{(2)}(0))$ \cite{ollivier_hong_2020}. This gives an upper bound to the HOM visibility taking the finite $g^{\left(2\right)}\left(0\right)$ into account. Before conversion, $\rm M_s$ is equal to $0.95\,(1)$. After conversion we find the raw visibility and corrected indistinguishability to be $0.60\,(1)$ and $0.67\,(2)$ respectively. 
The results of lifetime, HBT and HOM measurements are summarised in Table ~\ref{tab:table1}.
The reduced interference visibility originates in spectral instability introduced by fluctuating power in multiple longitudinal modes of the seed laser. The line width of the seed laser is around 4 GHz with a free spectral range estimated to be 177 MHz, corresponding to $\sim 22$ modes. 
Despite this, we show that our QFC setup indeed preserves the coherence of single photons as measured from non-classical two-photon interference which can be improved by increased control over the cavity dispersion and active stabilization of the cavity length. 

\begin{table}[tb]
\centering
\begin{ruledtabular}
\begin{tabular}{ c c c}
\hline
\textrm{}&
\textrm{942 nm} &
\textrm{1550 nm}\\
\hline
Lifetime $T_1\,(\mathrm{ns}$) & 0.2622 (2) & 0.2621 (2)\\
Resonant count rate (kHz) & 1,460 (40) & 456 (14) \\
Off-resonant count rate (kHz) & 1,850 (50) & 856 (18)\\
Off-resonant $g^{(2)}(0)$  & 0.045 (0) & 0.051 (1)\\
Resonant $g^{(2)}(0)$  & 0.040 (0) & 0.043 (1)\\
Resonant $\rm V_{HOM}$ & 0.88 (1) & 0.60 (1) \\
Resonant $M_s$ & 0.95 (1) & 0.67 (2)\\
\hline
\end{tabular}
\end{ruledtabular}
\caption{\label{tab:table1}%
Summary of the lifetime, count rate, $g^{(2)}(0)$ and indistinguishability $\rm V_{HOM}$ for converted and unconverted photons. 
Values are obtained from measurement results, illustrated in Fig.~\ref{fig:qd_fig_2}.
The corrected photon indistinguishability $M_S$ is estimated based on the measured $g^{(2)}(0)$  and uncorrected $\rm V_{HOM}$ \cite{ollivier_hong_2020}.
The error, given by the standard deviation from the fit, is included in brackets.
}
\end{table}

Modest improvements to the current QFC system will allow us to improve the converted two-photon interference visibility to equal the unconverted visibility.  The external conversion efficiency could be further improved with lower-loss filtering and improved mode matching between the single-mode fibre and the waveguide mode. 
We believe that this source will find applications in fibre-based quantum communication where a source of bright and highly pure single photons in the C-band is required. This can lead to demonstrations of various quantum communication protocols including measurement-device-independent quantum key distribution, teleportation and entanglement swapping between distant quantum nodes.

\begin{acknowledgments}
We acknowledge Pierre M. Petroff for his contribution to the sample design and fabrication. This work was supported by the EPSRC (Grants No. EP/L015110/1, No. EP/M013472/1, No. EP/P029892/1, EP/N002962/1 and  EP/T001011/1), the ERC (Grant No. 725920), and the EU Horizon 2020 research and innovation program under Grant Agreement No. 820423. B. D. G. thanks the Royal Society for a Wolfson Merit Award and the Royal Academy of Engineering for a Chair in Emerging Technology.
Y. M. acknowledges the support from Chongqing Research Program of Basic Research and Frontier Technology (No.cstc2016jcyjA0301). The authors in K.I.S.T. acknowledge the support from the KIST institutional program, the program of quantum sensor core technology through IITP and the IITP grant funded by the Korea government(MSIT) (No. 20190004340011001).
\end{acknowledgments}

\section*{References}

\bibliography{biblio}

\end{document}